\documentclass[aps,prd,onecolumn,groupedaddress,showpacs,nofootinbib,amssymb]{revtex4}
\usepackage{graphicx,bm,color}
\usepackage{amsmath}
\usepackage{amssymb}
\usepackage{amsfonts}
\usepackage{braket}
\usepackage{bm}

\makeatletter
\renewcommand{\theequation}{\arabic{section}.\arabic{equation}}
\@addtoreset{equation}{section}
\makeatother

\newcommand{\be}{\begin{equation}}
\newcommand{\ee}{\end{equation}}
\newcommand{\bea}{\begin{eqnarray}}
\newcommand{\eea}{\end{eqnarray}}
\newcommand{\beaa}{\begin{eqnarray*}}
\newcommand{\eeaa}{\end{eqnarray*}}

\newcommand{\e}{\mathrm{e}}

\newcommand{\f}[2]{\frac{#1}{#2}}

\newcommand{\Rt}{\right}
\newcommand{\Lt}{\left}
\newcommand{\p}{\partial}
\newcommand{\n}{\nonumber}

\allowdisplaybreaks[4] 

\begin{document}

\title{On relationship of gauge transformation with Wigner's little group}

\author{
Hiroshi Yoda
}

\affiliation{
 Department of Physics, Nagoya University, Nagoya 464-8602, Japan \\
}

\begin{abstract}
Wigner's little group of a massless particle is $ISO(2)$ which contains rotation and two translations.
As well]known, eigenvalues of the rotation are helicity.
On the other hand, by S. Weinberg et al., it has been shown that two translations generate abelian gauge transformation by acting on polarization vectors.
In this paper, we include unphysical modes and show abelian case result can be generalized to the case of non-abelian gauge transformation.
By including the unphysical modes, we obtain Nakanishi-Lautrup physical state condition from the requirement of unitarity of the transformation.
As a result, non-abelian gauge transformation is realized as the translation of the little group which acts on gauge group.
We also obtain similar results for any spacetime dimensions.
\end{abstract}
\maketitle

\section*{INTRODUCTION}
In the context of relativistic quantum mechanics, a free single-particle state is defined as unitary irreducible representation of the Poincar\'{e} group.
E. Wigner\cite{Wigner} has investigated that such representation is classified by the mass, spatial momenta and labels of the {\it little group}. 
To begin with, we review such classification briefly.

In particular, for massless particle the little group is $ISO(2)$, namely two dimensional Euclidean group which contains one rotation and two translations.
These operators act on field operators under the Lorentz transformation.
The rotation generate $U(1)$ phase transformation corresponding to eigenvalues called {\it helicity}.
On the other hand, two translations generate {\it abelian} gauge transformation by acting on transverse polarization vectors.
This result has been shown by S. Weinberg et al.\cite{LT-abelian} and also generalized for the case of spin-2 or higher-spin fields.
It is very interesting that these fact means abelian {{\it gauge symmetry is induced by only imposing Lorentz symmetry} for the Lagrangian of massless integer spin fields.

We review and generalize this interesting result.
First, we consider the case of including unphysical, namely, longitudinal and scalar modes.
And then, we show that this result can be generalized to the case of non-abelian gauge transformation and for any spacetime dimensions.

\section{A brief review of Wigner's little group}
In this section, we present representation of the Poincar\'{e} group and Wigner's little group.
A {\it relativistic free single-particle state} is defined as unitary irreducible representation of the Poincar\'{e} group which is generated by the Poincar\'{e} algebra;
\begin{align}
[P_\mu ,P_\nu ]=0\, ,\quad
[M_{\mu \nu},M_{\rho \sigma}]=\eta  _{\mu\rho}M_{\nu\sigma}+\eta  _{\nu\sigma}M_{\mu\rho}-\eta  _{\nu\rho}M_{\mu\sigma}-\eta  _{\mu\sigma}M_{\nu\rho}\, ,\quad
[M_{\mu \nu},P_\rho ]=\eta _{\mu\rho}P_\nu-\eta _{\nu\rho}P_\mu\, .
\label{Poincare}
\end{align}
There are two Casimir operators of the Poincar\'{e} algebra;
\begin{align}
P^2:=P^\mu P_\mu\, ,\quad W^2:=W^\mu W_\mu\, ,
\end{align}
where,
\begin{align}
 W_\mu :=-\f{1}{2}\varepsilon _{\mu \nu \rho \sigma}P^\nu M^{\rho \sigma}\, .
\end{align}
Therefore, we can classify the representation by the square of mass $p^2$; eigenvalues of $P^2$.
Furthermore, $W_\mu$ generates several different algebras; called {\it little group}, depnding on $p^\mu$ as in the following table: 
\begin{center}
\begin{tabular}{cc|c}
\quad
eigenvalues of $P^2$ && little group\\
\hline
timelike
&
$p^2>0$
\quad
&
$SO(3)$
\\
null
&
$p^2=0$
\quad
&\quad
$ISO(2)$
(2-dimensional Euclidean group)
\\
spacelike
&
$p^2<0$
\quad
&
$SO(2,1)$
\\
no particles
&
$p^\mu=0$
\quad
&
$SO(3,1)$
\end{tabular}
\end{center}
Furthermore, because of $[P_\mu ,W_\nu ]=0$, labels of the representation of the little group are independent of $p^\mu$.
Consequently, we can classify completely the representation of Poincar\'{e} group by mass, spatial momenta and labels of the little group.

Next, we consider specific transformation law of such relativistic free single-particle states.
Since the little group transformation does not change momenta, we can choose arbitrary reference frame.
First, we consider massive ($p^2>0$) particles and choose the static frame $k^\mu=(m,0,0,0)$.
Hence, we introduce the little group transformation matrix $W$ as follows.
\begin{align}
W^\mu _{\ \nu}(\Lambda , p )k^\nu =k^\mu \,.
\end{align} 
This matrix can be rewritten by using two types of the Lorentz transformation;
the generic transformation $\Lambda$ and the standard transformation $L(p)$ which satisfies $L(p)^\mu _{\ \nu}k^\nu=p^\mu$.
Then, we can find
\begin{align}
W(\Lambda , p )=L^{-1}(\Lambda p)\Lambda L(p) \,.
\label{W}
\end{align}
By using (\ref{W}), we obtain the transformation law of free single-particle states $\Ket{{\bf p},\sigma }$.
Now, $\sigma$ is the index of internal symmetry, namely the little group $SO(3)$, called {\it spin}.
That is, 
\begin{align}
U(\Lambda )\Ket{{\bf p},\sigma }&=U(L(\Lambda p) )U(W(\Lambda , p ) )U^{-1}(L(p) )\Ket {{\bf p},\sigma } \n\\
&=U(L(\Lambda p) )U(W(\Lambda , p ) )\Ket {{\bf 0},\sigma } \n\\
&=D(\Lambda , p )_{\sigma \sigma '}\Ket {{\bf \Lambda p},\sigma '}\,.
\end{align}
$D(\Lambda , p )_{\sigma \sigma '}$ is some unitary representation of $SO(3)$ called {\it Wigner rotation}.
Consequently, the Lorentz transformation of free massive single-particle states is composed with both Lorentz boost of spatial momenta and the rotation of spin state.

After that, we consider massless ($p^2=0$) particles and choose the reference frame $k^\mu=(k,0,0,k)$.
We can also use the relation (\ref{W}).
However, the little group is $ISO(2)$;
\begin{align}
[J,T_1]=iT_2\,,
\quad
[J,T_2]=-iT_1\,,
\quad 
[T_1,T_2]=0\,.
\label{ISO(2)}
\end{align}
$J$ is generator of $SO(2)$ but $T_1$ and $T_2$ are generator of translation.
Therefore, states are labeled by one discrete index called {\it helicity} and two continuous indices.
Since it is necessary to consider up to two-valued representation of $SO(2)$, the helicity index take integer or half-integer same as ordinary spin index.
Moreover, in the current reference frame $J$ is the rotation generator of around z-axis.
So, indeed, helicity is projected angular momentum to z-axis and regard as similar degree of freedom to spin.  
On the other hands, there is no physical quantity which corresponds to continuous indices called {\it continuous spin}.
Thus, we define free single-particle states as only having the helicity index $h$; $\Ket {{\bf p},h}$
and action of $J$, $T_1$ and $T_2$ as follows.
\begin{align}
J\Ket {{\bf p},h}=h\Ket {{\bf p},h}\, ,\quad T_1\Ket {{\bf p},h}=T_2\Ket {{\bf p},h}=0\, .
\end{align}
Hence, we obtain 
\begin{align}
U(\Lambda )\Ket {{\bf p},h}=\e ^{i\theta (\Lambda ,p)h}\Ket{ {\bf \Lambda p},h}\, .
\label{Lt-ket}
\end{align}
Consequently, the Lorentz transformation of free massless single-particle states is composed with both Lorentz boost of spatial momenta and the $U(1)$ phase transformation differently from the case of massive particles.

\section{Abelian gauge transformation}
In this section, we consider Lorentz transformation of massless spin-1,2 and any integer spin field operators
and show $U(1)$ gauge transformation, linearized general coordinate transformation and higher-spin gauge transformation are induced by the the Lorentz transformation, respectively.
\subsection{Massless fields of integer spin}
We introduce creation and annihilation operators;
\begin{align}
\Ket {{\bf p},h}=\sqrt{2E_{{\bf p}}}a^\dagger_h({\bf p})\Ket{0}\,,\quad
[a_h({\bf p}),a^\dagger _j ({\bf q})]=\delta _{hj}\delta ^3({\bf p-q})\, ,
\label{OPphys}
\end{align}
and obtain the Lorentz transformation of these operators from (\ref{Lt-ket});
\begin{align}
U(\Lambda )a^\dagger _h ({\bf p})U^\dagger (\Lambda )= \sqrt{\f{E_{{\bf\Lambda p}}}{E_{{\bf p}}}}\e ^{i\theta h}a^\dagger _h ({\bf\Lambda p})\,,
\quad 
U(\Lambda )a_h({\bf p})U^\dagger (\Lambda )=\sqrt{\f{E_{{\bf\Lambda p}}}{E_{{\bf p}}}} \e ^{-i\theta h}a_h({ \bf \Lambda p})\, .
\label{Lt-a}
\end{align}

To construct Lorentz covariant operators, we define two polarization vectors in the reference frame $k^\mu=(k,0,0,k)$ as
\begin{align}
\epsilon ^\mu _{\pm 1}:=\f{1}{\sqrt{2}}(0,1,\pm i,0)\,,
\end{align}
and find
\begin{align}
k_\mu\epsilon ^\mu _{\pm 1}=(\epsilon ^\mu _{ \pm 1})^2=0\,,
\quad
\epsilon ^{*\mu }_{\pm 1}=\epsilon ^\mu _{\mp 1}\,,
\quad
\epsilon  ^*_{ \pm 1\mu}\epsilon ^\mu _{\pm 1}=1\,.
\label{epsilon}
\end{align}
Then, the little group $ISO(2)$ generators acting polarization vectors are
\begin{align}
(T_1)^\mu _{\ \nu} =
\begin{pmatrix}
0 & -i & 0 & 0 \\
-i & 0 & 0 & i \\
0 & 0 & 0 & 0\\
0 & -i & 0 & 0
\end{pmatrix}\,,
\quad
(T_2)^\mu _{\ \nu} =
\begin{pmatrix}
0 & 0 & -i & 0 \\
0 & 0 & 0 & 0 \\
-i & 0 & 0 & i\\
0 & 0 & -i & 0
\end{pmatrix}\,,
\quad
(J)^\mu _{\ \nu} =
\begin{pmatrix}
0 & 0 & 0 & 0 \\
0 & 0 & -i & 0 \\
0 & i & 0 & 0\\
0 & 0 & 0 & 0
\end{pmatrix}\,,
\end{align}
and we obtain 
\begin{align}
W^\mu _{\ \nu}\epsilon ^\nu _{\pm 1}=(\e ^{i\alpha _1 T_1+i\alpha _2 T_2+i\theta J})^\mu _{\ \nu}\epsilon ^\nu _{\pm 1}
\simeq (1\pm i\theta )\epsilon ^\mu _{\pm 1}+\f{\alpha _1\pm i\alpha _2 }{\sqrt{2}k}k^\mu\,.
\label{Wepsilon}
\end{align}
By using the standard Lorentz transformation $L(p)$, we define polarization vectors in general frame as
\begin{align}
\epsilon ^\mu _{\pm 1}({\bf p}):=L^\mu _\nu (p)\epsilon ^\nu _{\pm 1}\, ,
\label{epsilon(p)}
\end{align}
find equations similar to (\ref{epsilon}) by replacing $\epsilon ^\mu _{\pm 1}\rightarrow \epsilon ^\mu _{\pm 1}({\bf p})$ and $k^\mu\rightarrow p^\mu$,
and obtain
\begin{align}
(1\mp i\theta )\epsilon ^\mu _{\pm 1}({\bf p})
\simeq (\Lambda ^{-1})^\mu _{\ \nu}\epsilon ^\nu _{\pm 1}({\bf \Lambda p})+\f{\alpha _1 \pm i\alpha _2}{\sqrt{2}k}p^\mu
\label{Lt-epsilon}
\end{align}
from (\ref{W}) and (\ref{Wepsilon}).

\subsubsection{Spin-1 (Electromagnetic field)}
We introduce the real vector field operator as follows.
\begin{align}
A^\mu (x):=\int \f{d^3{ \bf p}}{(2\pi )^\f{3}{2}\sqrt{2E_{{\bf p}}}}\sum_{h=\pm 1}\Lt[ \epsilon ^\mu _h ({\bf p}) a_h({\bf p}) \e ^{-ipx}
+\epsilon ^{ *\mu} _h ({\bf p}) a^\dagger _h({\bf p}) \e ^{ipx}\Rt]\, .
\label{A^mu} 
\end{align}
We should note that  $A^\mu (x)$ is fixed to the Lorenz gauge $\p_\mu A^\mu (x)=0$.
By using (\ref{Lt-a}) and (\ref{Lt-epsilon}), we obtain infinite small Lorentz transformation of  (\ref{A^mu}) is 
\begin{align}
U(\Lambda )A^\mu (x)U^\dagger (\Lambda )
&\simeq \int \f{d^3{ \bf p}\sqrt{2E_{{\bf\Lambda p}}}}{(2\pi )^\f{3}{2}2E_{{\bf p}}}\sum_{h=\pm 1}
\Lt[ (1-i\theta h)\epsilon ^\mu _h ({\bf p}) a_h({\bf \Lambda p}) \e ^{-ipx}+ (1+i\theta h)\epsilon ^{*\mu} _h ({\bf p}) a^\dagger _h({\bf \Lambda p}) \e ^{ipx}\Rt] \n\\
&=  \int  \f{d^3{ \bf p}\sqrt{2E_{{\bf\Lambda p}}}}{(2\pi )^\f{3}{2}2E_{{\bf p}}}\sum_{h=\pm 1}
\Lt[ \Lt( (\Lambda ^{-1})^\mu _{\ \nu}\epsilon ^\nu _h ({\bf\Lambda p})  
+\f{\alpha _1+ ih\alpha _2 }{\sqrt{2}k}p^\nu \Rt)a_h({\bf \Lambda p})\e ^{-ipx} \Rt.\n\\
&\hspace{15em}\Lt.
+ \Lt((\Lambda ^{-1})^\mu _{\ \nu}\epsilon ^{*\nu} _h ({\bf\Lambda p})
+\f{\alpha _1- ih\alpha _2 }{\sqrt{2}k}p^\nu\Rt) a^\dagger _h({\bf \Lambda p}) \e ^{ipx}\Rt] \n\\
&=(\Lambda ^{-1})^\mu _{\ \nu}\Lt( A^\nu (\Lambda x) +\p ^\nu \Theta  (\Lambda x ,\alpha _1,\alpha _2)\Rt)\, .
\label{Lt-A}
\end{align}
Now, we define the local operator
\begin{align}
\Theta  (x,\alpha _1,\alpha _2):=\f{i}{\sqrt{2}k}\int \f{d^3{ \bf p}}{(2\pi )^\f{3}{2}\sqrt{2E_{{\bf p}}}}\sum_{h=\pm 1}
\Lt[\alpha _1 ( a_h({\bf p}) \e ^{-ipx}-a^\dagger _h({\bf p})\e ^{ipx}) +ih\alpha _2( a_h({\bf p}) \e ^{-ipx}+a^\dagger _h({\bf p})\e ^{ipx})\Rt]\, .
\label{Theta}
\end{align}
The important point to be noted is that $\Theta  (x,\alpha _1,\alpha _2)$ is hermitian. 
Consequently, we find that abelian gauge transformation of gauge field is induced by the Lorentz transformation,
and particularly, for factor of $\alpha _1$ and $\alpha _2$,  generated by two translations of $ISO(2)$.
Lastly, we should note that the gauge fixing condition $\p_\mu A^\mu (x)=0$ is invariant in this gauge transformation because of the Lorentz invariance
\footnote{
Also we can check this invariance by specific calculation using $p^2=0$.
}
of the gauge fixing condition.
This result is consistent with gauge invariance of physical states.

\subsubsection{Spin-2 (Linearized gravity)}
We can obtain gauge transformation of linearized gravity from the Lorentz transformation by the similar way to spin-1 field.
First, we introduce creation and annihilation operators of spin-2 field by taking symmetric tensor products of spin-1 operators;
\begin{align}
a_{2h}({\bf p}):=a_h({\bf p})\otimes a_h({\bf p})\, .
\end{align}
Hence, we can find 
\begin{align}
U(\Lambda )a_{2h}({\bf p})U^\dagger (\Lambda )
= \sqrt{\f{E_{{\bf\Lambda p}}}{E_{{\bf p}}}}\e ^{-2i\theta h}a_{2h}({\bf \Lambda p}) \,.
\label{a_2h}
\end{align}
Then, we also introduce two polarization tensors;
\begin{align}
\epsilon ^{\mu\nu} _{\pm 1}({\bf p}):=\epsilon ^\mu _{\pm 1}({\bf p})\otimes \epsilon ^\nu _{\pm 1}({\bf p})\,.
\label{epsilon_2h}
\end{align}
By using (\ref{a_2h}) and (\ref{epsilon_2h}), we define the real rank-2 tensor field as follows. 
\begin{align}
h^{\mu\nu} (x):=
\int \f{d^3{ \bf p}}{(2\pi )^\f{3}{2}\sqrt{2E_{{\bf p}}}}\sum_{h=\pm 1}\Lt[ \epsilon ^{\mu\nu} _h ({\bf p}) a_{2h}({\bf p}) \e ^{-ipx}
+\epsilon ^{*\mu\nu} _h ({\bf p}) a^\dagger _{2h}({\bf p}) \e ^{ipx}\Rt] \,,
\label{h}
\end{align}
Obviously, $h^{\mu\nu} (x)$ is a symmetric tensor
and satisfy traceless transverse gauge condition from (\ref{epsilon});
\begin{align}
\p_\mu h^{\mu\nu} (x)=0,\quad
\eta_{\mu\nu}h^{\mu\nu} (x)=0 \,.
\end{align}

We can find Lorentz transformation of $h^{\mu\nu} (x)$ is
\begin{align}
U(\Lambda )h^{\mu\nu}(x)U^\dagger (\Lambda )
\simeq (\Lambda ^{-1})^\mu _{\ \rho}(\Lambda ^{-1})^\nu _{\ \sigma}
\Lt(h^{\rho \sigma } (\Lambda x) +\p ^\rho \xi  ^\sigma  (\Lambda x)+\p ^\sigma \xi  ^\rho  (\Lambda x)\Rt)\, ,
\label{Lt-h}
\end{align}
where,
\begin{multline}
\xi  ^\mu  (x):=\f{i}{\sqrt{2}k}\int \f{d^3{ \bf p}}{(2\pi )^\f{3}{2}\sqrt{2E_{{\bf p}}}}\sum_{h=\pm 1}
\Lt[\alpha _1(\epsilon ^\mu _h ({\bf p})a_{2h}({\bf q}) \e ^{-ipx}-\epsilon ^{*\mu }_h({\bf p})a^\dagger _{2h}({\bf p}) \e ^{ipx})
\Rt.\\ \Lt.
+ih\alpha _2(\epsilon ^\mu _h ({\bf p})a_{2h}({\bf p}) \e ^{-ipx}+\epsilon ^{*\mu }_h({\bf p})a^\dagger _{2h}({\bf p}) \e ^{ipx})\Rt]\, ,
\end{multline}
\begin{align}
\xi  ^{\dagger \mu}  (x)=\xi  ^\mu  (x)\, .
\end{align}
Consequently, we can regard $h^{\mu\nu} (x)$ as graviton and (\ref{Lt-h}) as linearized general coordinate transformation.

\subsubsection{Spin-n (Higher-spin gauge field)}
From generalization of the case of spin-2 field, we can obtain gauge transformation of free higher-spin gauge theory.
Free integer-spin massless field theory in 4-dimensional spacetimes has found by Fronsdal \cite{Fronsdal}
and known this theory has abelian gauge symmetry which is like generalization of linearized general coordinate transformation.

Similarly to the case of spin-2, we introduce creation and annihilation operators of spin-n field by taking symmetric tensor products of spin-1 operators;
\begin{align}
a_{nh}({\bf p}):=( a_h({\bf p}) )^{\otimes n}\, .
\label{a_nh}
\end{align}
Then, we also introduce two polarization tensors;
\begin{align}
\epsilon ^{\mu_1\cdots\mu_n} _{\pm 1}({\bf p})
:=\epsilon ^{(\mu_1} _{\pm 1}({\bf p})\otimes\cdots\otimes \epsilon ^{\mu_n )} _{\pm 1}({\bf p})\, .
\label{epsilon_nh}
\end{align}
$(\mu_1\cdots \mu_n)$ means to take totally-symmetric product.
By using (\ref{a_nh}) and (\ref{epsilon_nh}), we define the real totally-symmetric rank-n tensor field as follows. 
\begin{align}
\phi ^{\mu_1\cdots\mu_n} (x):=
\int \f{d^3{ \bf p}}{(2\pi )^\f{3}{2}\sqrt{2E_{{\bf p}}}}\sum_{h=\pm 1} \Lt[ \epsilon ^{\mu_1\cdots\mu_n} _{h} ({\bf p}) a_{nh}({\bf p}) \e ^{-ipx}
+\epsilon ^{*\mu_1\cdots\mu_n} _{h} ({\bf p}) a^\dagger _{nh}({\bf p}) \e ^{ipx} \Rt] \,.
\label{HS}
\end{align}
From (\ref{epsilon}), $\phi ^{\mu_1\cdots\mu_n} (x)$ satisfies the gauge fixing condition \cite{de Wit-Freedman};
\begin{align}
\p_\nu \phi ^{\nu\mu_2\cdots\mu_n} (x)=0\,,
\quad
\eta_{\nu_1 \nu_2 }\phi ^{\nu_1 \nu_2 \mu_3\cdots\mu_n}=0 \,.
\end{align}

We can also find Lorentz transformation of $\phi ^{\mu_1\cdots\mu_n} (x)$ is
\begin{align}
U(\Lambda )\phi ^{\mu_1\cdots\mu_n} (x)U^\dagger (\Lambda )
=(\Lambda ^{-1})^{\mu_1} _{\ \nu_1}\cdots(\Lambda ^{-1})^{\mu_n} _{\ \nu_n}
\Lt(\phi ^{\nu_1\cdots\nu_n}  (\Lambda x) +\p ^{(\nu_1} \xi  ^{\nu_2\cdots\nu_n)}   (\Lambda x)\Rt)\, ,
\label{Lt-HS}
\end{align}
where,
\begin{multline}
\xi  ^{\mu_1\cdots\mu_{n-1}}  (x):=\f{i}{\sqrt{2}k}\int \f{d^3{ \bf p}}{(2\pi )^\f{3}{2}\sqrt{2E_{{\bf p}}}}\sum_{h=\pm 1}
\Lt[ \alpha _1(\epsilon ^{\mu_1\cdots\mu_{n-1}} _{h}({\bf p})a_{nh}({\bf p}) \e ^{-ipx}-\epsilon ^{*\mu_1\cdots\mu_{n-1}}_{h}({\bf p})a^\dagger _{nh}({\bf p}) \e ^{ipx})
\Rt. \\ \Lt.
+ih\alpha _2(\epsilon ^{\mu_1\cdots\mu_{n-1}} _{h}({\bf p})a_{nh}({\bf p}) \e ^{-ipx}+\epsilon ^{*\mu_1\cdots\mu_{n-1}}_{h}({\bf p})a^\dagger _{nh}({\bf p}) \e ^{ipx})\Rt] \, ,
\end{multline}
\begin{align}
\xi  ^{\dagger \mu_1\cdots\mu_{n-1}}  (x)=\xi  ^{\mu_1\cdots\mu_{n-1}}  (x) \, ,
\quad
\eta_{\nu_1\nu_2}\xi  ^{\nu_1\nu_2\mu_3\cdots\mu_{n-1}}  (x)=0 \,.
\end{align}
We should note that this traceless condition of gauge parameter $\xi  ^{\dagger \mu_1\cdots\mu_{n-1}}  (x)$ has been required in \cite{Fronsdal}.
Consequently, we can regard $\phi ^{\mu_1\cdots\mu_n} (x)$ as higher-spin gauge field and (\ref{Lt-HS}) as linearized general coordinate transformation.

\subsection{Including unphysical modes and representation of the little group}
In previous section, we consider the massless vector field operator composed of only physical, namely, two transverse modes
In this section, on the other hand, we consider not only physical modes but also two unphysical modes, namely, longitudinal and scalar mode.

$A^\mu (x)$ has, apparently, four components, but only two degree of freedom are physical because of gauge fixing $\p_\mu A^\mu (x)=0$. 
Now, we consider remaining longitudinal and scalar mode in reference frame
\begin{align}
\epsilon ^\mu _L:=\f{1}{\sqrt{2}ik}(k,0,0,k)\, ,
\quad
\epsilon ^\mu _S:=\f{i}{\sqrt{2}k}(k,0,0,-k)\,,
\end{align}
and find
\begin{align}
k_\mu\epsilon ^\mu _L=(\epsilon ^\mu  _L)^2=(\epsilon ^\mu _S)^2=0\,,
\quad
k_\mu\epsilon ^\mu _S=\sqrt{2}ik\, ,
\quad
\epsilon _{L\mu}\epsilon ^\mu _S=1\,,
\label{epsilonLS}
\end{align}
\begin{align}
W^\mu _{\ \nu}\epsilon ^\nu _L=\epsilon ^\mu _L\,,
\quad
W^\mu _{\ \nu}\epsilon ^\nu _S
\simeq\epsilon ^\mu _S+\alpha _1(\epsilon ^\mu _{+1}+\epsilon ^\mu _{-1})+\alpha _2\f{1}{i}(\epsilon ^\mu _{+1}-\epsilon ^\mu _{-1})\,.
\label{WepsilonLS}
\end{align}
Moreover, we should note that physical and unphysical modes satisfy the completeness relation;
\begin{align}
\sum _{h=\pm 1}\epsilon ^{*\mu} _h\epsilon ^\nu _h+\epsilon ^{*\mu} _L\epsilon ^\nu _S+\epsilon ^{*\mu} _S\epsilon ^\nu _L=-\eta^{\mu\nu}\,.
\label{completeness}
\end{align}
Similarly to (\ref{epsilon(p)}), we define unphysical mode in general frame.
Then, we obtain similar relations to (\ref{epsilonLS}) and (\ref{completeness}), and
\begin{align}
\epsilon ^\nu _L({\bf p})=(\Lambda ^{-1})^\mu _{\ \nu} \epsilon ^\nu _L({\bf \Lambda p})\,, 
\label{WepsilonL(p)}
\end{align}
\begin{align} 
\epsilon ^\nu _S({\bf p})
\simeq\Lt( \Lambda ^{-1}\Rt) ^\mu _{\ \nu} \Lt( \epsilon ^\nu _S({\bf \Lambda p})+i\alpha _1\omega ^\nu ({\bf \Lambda p})+\alpha _2\tilde{\omega} ^\nu ({\bf \Lambda p})\Rt)\,, 
\label{WepsilonS(p)}
\end{align}
where,
\begin{align}
\omega  ^\mu ({\bf p}):= \epsilon ^\mu _{+1}({\bf p})+\epsilon ^\mu _{-1}({\bf p})\,,
\quad
\tilde{\omega } ^\mu ({\bf p}):=\f{1}{i}\Lt(\epsilon ^\mu _{+1}({\bf p})-\epsilon ^\mu _{-1}({\bf p})\Rt)\,.
\end{align}
Furthermore, we introduce creation and annihilation operators of unphysical modes as
\footnote{By using (\ref{OPphys}), (\ref{OPunphys}) and (\ref{completeness}), and defining $a^\mu({\bf p}):=\sum _{h=\pm 1,L,S}\epsilon ^\mu _h({\bf p}) a_h({\bf p})$ 
we obtain $[a^\mu ({\bf p}),a^{\nu\dagger} ({\bf q})]=-\eta^{\mu\nu}\delta ^3({\bf p-q})$.
}
\begin{align}
[a_L({\bf p}),a^\dagger _L({\bf q})]=[a_S({\bf p}),a^\dagger _S({\bf q})]=0\, ,\quad
[a_L({\bf p}),a^\dagger _S ({\bf q})]=\delta ^3({\bf p-q})\, .
\label{OPunphys}
\end{align}
Then, we introduce the real vector field operator including unphysical modes as follows.
\begin{align}
A^\mu (x):=\int \f{d^3{ \bf p}}{(2\pi )^\f{3}{2}\sqrt{2E_{{\bf p}}}}\sum_{h=\pm 1,L,S}\Lt[ \epsilon ^\mu _h ({\bf p}) a_h({\bf p}) \e ^{-ipx}
+\epsilon ^{ *\mu} _h ({\bf p}) a^\dagger _h({\bf p}) \e ^{ipx}\Rt]\, .
\label{A^mu unpys} 
\end{align}
We should note that $A^\mu (x)$ is fixed by general Lorentz invariant gauge condition
\begin{align}
\p_\mu A^\mu (x)=\sqrt{2}ik\int \f{d^3{ \bf p}}{(2\pi )^\f{3}{2}\sqrt{2E_{{\bf p}}}}\Lt[a_S({\bf p}) \e ^{-ipx}+a^\dagger _S({\bf p}) \e ^{ipx}\Rt]\, ,
\label{GF-A unphys}
\end{align}
and the right hand side of (\ref{GF-A unphys}) is the Nakanishi-Lautrap auxiliary field $B(x)$\cite{NL}.

We consider Lorentz transformation of (\ref{A^mu unpys}),
but we do not determine similarly the Lorentz transformation of these operators to the procedures of section I
because $a^\dagger _L({\bf q})$ and $a^\dagger _S({\bf q})$ are creation operators corresponded to unphysical states.
So, we set these transformation for not changing gauge fixing condition (\ref{GF-A unphys}) because of the Lorentz invariance.
Furthermore, by considering (\ref{WepsilonL(p)}) and (\ref{WepsilonS(p)}), we assume that little group should not act unphysical states.
Then, we set
\begin{align}
U(\Lambda )a _L ({\bf p})U^\dagger (\Lambda )= \sqrt{\f{E_{{\bf\Lambda p}}}{E_{{\bf p}}}}a _L ({\bf\Lambda p})\,,
\quad 
U(\Lambda )a_S({\bf p})U^\dagger (\Lambda )=\sqrt{\f{E_{{\bf\Lambda p}}}{E_{{\bf p}}}} a_S({ \bf \Lambda p})\, .
\label{Lt-a unphys0}
\end{align}
By using (\ref{Lt-a}), (\ref{Lt-epsilon}), (\ref{WepsilonL(p)}), (\ref{WepsilonS(p)}) and (\ref{Lt-a unphys0}) consequently, we obtain
\begin{align}
U(\Lambda )A^\mu (x)U^\dagger (\Lambda )
&\simeq (\Lambda ^{-1})^\mu _{\ \nu}\Lt( A^\nu (\Lambda x) +\p ^\nu \Theta  (\Lambda x ,\alpha _1,\alpha _2)+ \Omega ^\nu (\Lambda x, \alpha _1,\alpha _2) \Rt)\, ,
\label{Lt-A unphys}
\end{align}
where,
\begin{align}
\Omega ^\mu (x, \alpha _1,\alpha _2):=i\int \f{d^3{ \bf p}}{(2\pi )^\f{3}{2}\sqrt{2E_{{\bf p}}}}
\Lt[\Lt(\alpha _1\omega ^\mu ({\bf  p})+\alpha _2\tilde{\omega} ^\mu ({\bf  p})\Rt) a _S({\bf p}) \e ^{-ipx}
-\Lt( \alpha _1\omega ^{*\mu} ({\bf  p})+\alpha _2\tilde{\omega} ^{*\mu} ({\bf  p})\Rt)a^\dagger _S({\bf p}) \e ^{ipx}\Rt]\, .
\end{align}
It is obvious that (\ref{Lt-A unphys}) is not gauge transformation.
So, we improve the transformation law of physical modes as follows
\begin{align}
\tilde{U}(\Lambda )a_h({\bf p})\tilde{U}^{-1} (\Lambda )
\simeq \sqrt{\f{E_{{\bf\Lambda p}}}{E_{{\bf p}}}} [(1-i\theta h) a_h({ \bf \Lambda p})
-i(\alpha _1-ih\alpha _2)a _S ({ \bf \Lambda p})]\, ,
\n
\end{align}
\begin{align}
\tilde{U} (\Lambda )a^\dagger _h ({\bf p})\tilde{U}^{-1}(\Lambda )
\simeq \sqrt{\f{E_{{\bf\Lambda p}}}{E_{{\bf p}}}}[(1+i\theta h) a^\dagger _h ({\bf\Lambda p})
+i(\alpha _1+ih\alpha _2)a^\dagger _S ({ \bf \Lambda p})]\,,
\label{Lt-a unphys}
\end{align}
and can find
\begin{multline}
\tilde{U}(\Lambda ' )\tilde{U}(\Lambda )a _h ({\bf p})\tilde{U}^{-1} (\Lambda )\tilde{U}^{-1} (\Lambda ' ) 
\\
\simeq \sqrt{\f{E_{{\bf\Lambda '\Lambda p}}}{E_{{\bf p}}}} [(1-i(\theta '+\theta )h) a_h({ \bf \Lambda '\Lambda p})
-i((\alpha _1'+\alpha _1)-ih(\alpha _2'+\alpha _2))a _S ({ \bf \Lambda '\Lambda p})]\, .
\end{multline}
Then, we obtain
\begin{align}
\tilde{U}(\Lambda )A^\mu (x)\tilde{U}^{-1} (\Lambda )
&\simeq (\Lambda ^{-1})^\mu _{\ \nu}\Lt( A^\nu (\Lambda x) +\p ^\nu \Theta  (\Lambda x ,\alpha _1,\alpha _2) \Rt)\, .
\label{Lt tilde-A unphysF}
\end{align}
However, (\ref{Lt-a unphys}) brakes the commutation relation $[a_L({\bf p}),a^\dagger _h({\bf q})]=0$.
Accordingly, we should also improve the transformation law of the longitudinal mode as follows.
\begin{align}
\tilde{U}(\Lambda )a_L({\bf p})\tilde{U}^{-1} (\Lambda )
\simeq \sqrt{\f{E_{{\bf\Lambda p}}}{E_{{\bf p}}}} [a_L({ \bf \Lambda p})
-i\sum_{h=\pm 1}(\alpha _1+ih\alpha _2)a _h ({ \bf \Lambda p})]\, ,
\label{Lt-aL}
\end{align}
\begin{align}
\tilde{U} (\Lambda )a^\dagger _L ({\bf p})\tilde{U}^{-1}(\Lambda )
\simeq \sqrt{\f{E_{{\bf\Lambda p}}}{E_{{\bf p}}}}[ a^\dagger _L ({\bf\Lambda p})
+i\sum_{h=\pm 1}(\alpha _1-ih\alpha _2)a^\dagger _h ({ \bf \Lambda p})]\,,
\label{Lt-aL dagger}
\end{align}
In fact, additional terms of (\ref{Lt-aL}) and (\ref{Lt-aL dagger}) generate inverse gauge transformation $-\p ^\mu \Theta $.
Consequently, we obtain
\begin{align}
\tilde{U}(\Lambda )A^\mu (x)\tilde{U}^{-1} (\Lambda )
= (\Lambda ^{-1})^\mu _{\ \nu}A^\nu (\Lambda x)\, .
\label{Lt tilde-A unphys}
\end{align}

The important point to note is that additional terms of the improved transformation which are proportional to $\alpha_1$ and $\alpha_2$ are representation of two translations of $ISO(2)$.
We define hermitian operators $T_1$, $T_2$ and $J$ as follows.
\begin{align}
[T_1,a _h({ \bf p})]=-a _S ({ \bf p})\,,
\quad
[T_2,a _h({ \bf p})]=iha _S ({ \bf p})\,,
\quad
[J,a _h({ \bf p})]=-ha _h({ \bf p})\,,
\n\\
[T_1,a^\dagger _h({ \bf p})]=a _S^\dagger ({ \bf p})\,,
\quad
[T_2,a^\dagger _h({ \bf p})]=iha _S^\dagger ({ \bf p})\,,
\quad
[J,a^\dagger _h({ \bf p})]=ha^\dagger _h({ \bf p})\,,
\label{T_1,T_2,J a}
\end{align}
\begin{align}
[T_1,a _L({ \bf p})]=-\sum_{h=\pm 1}a _h ({ \bf p})\,,
\quad
[T_2,a _L({ \bf p})]=-i\sum_{h=\pm 1}ha _h ({ \bf p})\,,
\n\\
[T_1,a^\dagger  _L({ \bf p})]=\sum_{h=\pm 1}a^\dagger  _h ({ \bf p})\,,
\quad
[T_2,a^\dagger  _L({ \bf p})]=-i\sum_{h=\pm 1}ha^\dagger  _h ({ \bf p})\,.
\label{T_1,T_2 aL}
\end{align}
Then, we obtain
\begin{align}
[[J,T_1],a _h({ \bf p})]=i[T_2,a _h({ \bf p})]\,,
\quad
[[J,T_2],a _h({ \bf p})]=-i[T_1,a _h({ \bf p})]\,,
\quad
[[T_1,T_2],a _h({ \bf p})]=0\,,
\end{align}
and similar relations for $a_L ({\bf p})$, $a^\dagger _h ({\bf p})$ and $a^\dagger_L ({\bf p})$,
while all commutators of $a _S ({\bf p})$ and $a^\dagger _S ({\bf p})$ vanish.
Namely, $T_1$ , $T_2$ and $J$ satisfy the algebra of $ISO(2)$ (\ref{ISO(2)}).

In addition, we can write down explicitly these $ISO(2)$ generators by using creation and annihilation operators as  follows.
\begin{align}
J=\int \!\! d^3{ \bf p} \sum _{h=\pm 1} h a^\dagger _h({ \bf p}) a _h({ \bf p})
\end{align}
\begin{align}
T_A=\sqrt{2}i\int \!\! d^3{ \bf p}\Lt(\tilde{a}^\dagger _A({ \bf p})a _S ({ \bf p}) -a _S^\dagger ({ \bf p}) \tilde{a} _A ({ \bf p})\Rt)
\end{align}
where,
\begin{align}
\tilde{a} _1:=\f{i}{\sqrt{2}}\sum_{h=\pm 1}a _h ({ \bf p})\, ,
\quad
\tilde{a} _2:=-\f{1}{\sqrt{2}}\sum_{h=\pm 1}ha _h ({ \bf p})\, ,
\end{align}
\begin{align}
[\tilde{a}_A({\bf p}),\tilde{a}_B ({\bf q})]=[\tilde{a}_A^\dagger ({\bf p}),\tilde{a}^\dagger _B ({\bf q})]=0\, ,
\quad
[\tilde{a}_A({\bf p}),\tilde{a}^\dagger _B ({\bf q})]=\delta _{AB}\delta ^3({\bf p-q})\, .
\end{align}
Furthermore, these operators vanish acting on vacuum;
\begin{align}
J\Ket{0}=T_A\Ket{0}=0\,.
\end{align}
This result are consistent with Lorentz invariance of vacuum.

Finally, we should note it is obvious that $\tilde{U} (\Lambda )$ does not act as the unitary operator on the Fock space despite hermiticity of $T_1$ , $T_2$ and $J$.
Unitarity of physical modes is broken by excited states of longitudinal mode because of (\ref{T_1,T_2 aL}).  
So, we should restrict the Fock space to the physical subspace which do not contain longitudinal mode such as,
\begin{align}
\{ \Ket{phys}\ ;\ a_S({ \bf p})\Ket{phys}=0\}\,.
\label{phys}
\end{align}
On the other hands, this physical subspace still contain excited states of scaler mode.
Since (\ref{T_1,T_2,J a}), unitarity of physical modes would be broken by existence of such states.
However, there is no such problem because these states are zero norm.
Consequently, continuous spin states generated by $T_A$ are unphysical or zero norm state. 
Furthermore, if we take expectation values of (\ref{GF-A unphys}) and (\ref{Lt-A unphys}) with $\Ket{phys}$, we obtain $\p_\mu A^\mu=0$ and gauge transformation (\ref{Lt-A}).
Moreover, since $a_S({ \bf p})$ is the annihilation operator of $B(x)$, the condition (\ref{phys}) is identical with the case of Nakanishi-Lautrup formalism\cite{NL}.

\section{Non-abelian gauge transformation}
In this section, we generalize the previous result that ``abelian gauge transformation is induced by the Lorentz transformation" to the case of non-abelian gauge group.

We introduce internal degree of freedom to massless spin-1 field.
Single-particle states degenerate into multiplet of gauge group $G$.
Now, we assume that $G$ is the compact semisimple Lie group and massless spin-1 field is the adjoint representation of $G$.
So, we should replace creation and annihilation operators so that $a_h ({\bf p})\rightarrow a^a_h ({\bf p})$
and commutation relations to
\begin{align}
[a^a_h({\bf p}),a^{b} _j ({\bf q})]=[a^a_h({\bf p\dagger}),a^{b\dagger} _j ({\bf q})]=0\, ,
\quad
[a^a_h({\bf p}),a^{b\dagger} _j ({\bf q})]=\delta ^{ab}\delta _{hj}\delta ^3({\bf p-q})\, .
\label{OP N-A}
\end{align}
Then, from the results of section II, we obtain abelian gauge transformation
\begin{align}
U(\Lambda )A^{\mu a} (x)U^\dagger (\Lambda )
&\simeq (\Lambda ^{-1})^\mu _{\ \nu}\Lt( A^{\nu  a}(\Lambda x) +\p ^\nu \Theta ^a(\Lambda x ,\alpha _1,\alpha _2) \Rt)\, .
\end{align}
This results is natural since we have consider the Lorentz transformation of {\it free} single-particle states.
In other worlds, self interaction terms of non-abelian gauge field should be ignored in our formulation.

Nevertheless, we can reproduce non-abelian gauge transformation from the {\it deformed Lorentz transformation} as follows.
\begin{multline}
U(\Lambda ,x)a^a_h({\bf p})U^\dagger (\Lambda ,x)
\simeq  \sqrt{\f{E_{{\bf\Lambda p}}}{E_{{\bf p}}}}\Lt[ (1-i\theta h) a^a_h({ \bf \Lambda p})
\Rt.\\ \Lt.
-g\Lt\{f^{abc}\Theta ^b (\Lambda x,\alpha _1,\alpha _2) a^c _h ({\bf\Lambda p})
-[a^a _h({ \bf \Lambda p}), f^{bcd}\Theta ^b (\Lambda x,\alpha _1,\alpha _2) ]N^{cd}
\Rt\}\Rt]\, ,
\label{Lt aa}
\end{multline}
\begin{multline}
U(\Lambda  ,x)a^{a\dagger}_h({\bf p})U^\dagger  (\Lambda  ,x)
\simeq  \sqrt{\f{E_{{\bf\Lambda p}}}{E_{{\bf p}}}}\Lt[(1+i\theta h) a^{a\dagger}_h({ \bf \Lambda p})
\Rt.\\ \Lt.
-g\Lt\{f^{abc}\Theta ^b (\Lambda x,\alpha _1,\alpha _2) a^c _h ({\bf\Lambda p})
-[a^{a\dagger}_h({ \bf \Lambda p}), f^{bcd}\Theta ^b (\Lambda x,\alpha _1,\alpha _2) ]N^{cd}
\Rt\}\Rt]\, .
\label{Lt aa dagger}
\end{multline}
and can find
\begin{multline}
U(\Lambda ' ,x)U(\Lambda ,x)a _h ({\bf p})U^\dagger  (\Lambda ,x)U^\dagger  (\Lambda ' ,x) 
\\
\simeq \sqrt{\f{E_{{\bf\Lambda '\Lambda p}}}{E_{{\bf p}}}}
\Lt[(1-i(\theta '+\theta )h) a^a_h({ \bf \Lambda '\Lambda p})
-g\Lt\{f^{abc}\Theta ^b (\Lambda '\Lambda  x,\alpha _1'+\alpha _1,\alpha _2'+\alpha _2) a^c _h ({\bf \Lambda '\Lambda p})\Rt.\Rt.
\\\Lt.\Lt.
-[a^a _h({ \bf \Lambda '\Lambda p}), f^{bcd}\Theta ^b (\Lambda '\Lambda  x,\alpha _1'+\alpha _1,\alpha _2'+\alpha _2) ]N^{cd}
\Rt\}\Rt]
\, .
\end{multline}
Now, 
\begin{align}
N^{ab}:=\int \!\! d^3{\bf p} \!\!  \sum _{h=\pm 1} \!\!  a^{a\dagger}_h ({\bf p}) a^b_h ({\bf p})
=\int \!\! d^3{\bf p} \!\!  \sum _{A=1,2} \!\!  \tilde{a}^{a\dagger}_A ({\bf p}) \tilde{a}^b_A ({\bf p}) \, ,
\end{align}
$g$ is gauge coupling constant and $f^{abc}$ is structure constant of $G$ which is totally anti-symmetric.
The last term of (\ref{Lt aa}) and (\ref{Lt aa dagger}) is needed for preserving commutation relations of $a^{a\dagger}_h$ and $a^a_h ({\bf p})$ (\ref{OP N-A}).

The important point to note is the transformation of  (\ref{Lt aa}) and (\ref{Lt aa dagger}) depend on coordinates $x$,
namely, the {\it local Lorentz transformation}.  
Then, we obtain
\begin{align}
U(\Lambda ,x)A^{\mu a}(x)U^\dagger (\Lambda  ,x)
\simeq (\Lambda ^{-1})^\mu _{\ \nu}\Lt(A^{\nu a} (\Lambda x) +\p ^\nu \Theta ^a (\Lambda x,\alpha _1,\alpha _2)  
+gf^{abc}A^{\nu b}(\Lambda x)\Theta ^c (\Lambda x,\alpha _1,\alpha _2)\Rt)\, .
\label{Lt-Aa}
\end{align}
Proportional terms to $N^{ab}$ vanish by the following reasons:
The form of these terms are
\begin{align}
f^{abc}N^{bc}\int \!\! \f{d^3{\bf p}}{(2\pi )^3 2E_{{\bf p}}} \Lt( \alpha_1 \omega ^\mu ({\bf p}) +\alpha_2 \tilde{\omega } ^\mu ({\bf p}) \Rt) \,.
\label{vanish}
\end{align}
Now,
$
\alpha_1 \omega ^\mu ({\bf p}) +\alpha_2 \tilde{\omega } ^\mu ({\bf p})
=L(p^0, {\bf p})( 0, \sqrt{2}\alpha_1 , \sqrt{2} i \alpha_2 ,0)
=L(p^0, -{\bf p})( 0, -\sqrt{2} \alpha_1 , - \sqrt{2} i \alpha_2 ,0) 
$.
Therefore, because of the invariance on ${\bf p}\rightarrow -{\bf p}$, (\ref{vanish}) vanish.

We should note that since both of $A^{\mu a}(x)$ and $\Theta ^a (x)$ are operators, there do not commute,
but $f^{abc}A^{\mu b}\Theta ^c=f^{abc}\Theta ^cA^{\mu b}$
because of the commutation relations (\ref{OP N-A}).
Consequently,  we find that non-abelian gauge transformation of gauge field is induced by the {\it local} transformation of  (\ref{Lt aa}) and (\ref{Lt aa dagger}).

Similarly to abelian transformation (\ref{Lt-a unphys}),
proportional terms to $\alpha_1$ and $\alpha_2$ are representation of two transformations of $ISO(2)$.
We define
\begin{align}
T_A^{ab}(x):=-f^{abc}\f{ig}{k}\int \f{d^3{ \bf p}}{(2\pi )^\f{3}{2}\sqrt{2E_{{ \bf p}}}}
\Lt[\tilde{a}^c_A({\bf p}) \e ^{-ipx}+\tilde{a}^{c\dagger} _A({\bf p}) \e ^{ipx}\Rt] \, ,
\end{align}
and find $(T_A^{ab}(x))^\dagger =-T_A^{ab}(x)$.
Then, we can rewrite (\ref{Lt aa}) and (\ref{Lt aa dagger}) for
\begin{align}
U (\Lambda ,x)a^a_h({\bf p})U^\dagger (\Lambda ,x)
\simeq  \sqrt{\f{E_{{\bf\Lambda p}}}{E_{{\bf p}}}} 
\Lt[(1-i\theta h) a^a_h({ \bf \Lambda p})
+\sum_{A=1,2} i\alpha _A \{ T_A^{ab}(x)a^b _h ({\bf\Lambda p})+[a^a _h ({\bf\Lambda p}),T_A^{bc}(x)]N^{bc} \}\Rt]\, ,
\label{Lt-aaT}
\end{align}
\begin{align}
U(\Lambda  ,x)a^{a\dagger}_h({\bf p})U^\dagger  (\Lambda  ,x)
\simeq  \sqrt{\f{E_{{\bf\Lambda p}}}{E_{{\bf p}}}} 
\Lt[(1+i\theta h) a^{a\dagger}_h({ \bf \Lambda p})
+\sum_{A=1,2} i\alpha _A \{ T_A^{ab}(x)a^{b\dagger} _h ({\bf\Lambda p})+[a^{a\dagger} _h ({\bf\Lambda p}),T_A^{bc}(x)]N^{bc} \}\Rt]\, .
\label{Lt-aaT dagger}
\end{align}
Thus, similarly to (\ref{T_1,T_2,J a}), we define hermitian operators $T_1(x)$, $T_2(x)$ and $J$ as follows.
\begin{align}
[T_A(x),a^a _h({ \bf p})]= T_A^{ab}(x)a^b _h ({\bf p})+[a^a _h ({\bf p}),T_A^{bc}(x)]N^{bc}\,,
\end{align}
\begin{align}
[J,a^a _h({ \bf p})]=-ha^a _h({ \bf p})\,.
\end{align}
By using 
$[J,T_1^{ab}]=iT_2^{ab}$, 
$[J,T_2^{ab}]=-iT_1^{ab}$,
$[a^a _h ,T_1^{bc}]=ih[a^a _h ,T_2^{bc}]$
and the commutation relation (\ref{OP N-A}), then, we obtain
\begin{align}
[[J,T_1(x)],a^a _h({ \bf p})]=i[T_2(x),a^a _h({ \bf p})]\,,
\quad
[[J,T_2(x)],a^a _h({ \bf p})]=-i[T_1(x),a^a _h({ \bf p})]\,,
\quad
[[T_1(x),T_2(x)],a^a _h({ \bf p})]=0\,,
\end{align}
and similar relations for $a^{a\dagger} _h ({\bf p})$.
In addition, we can also write down explicitly $T_A$ and $J$ by using the creation and annihilation operators as  follows
\begin{align}
T_A(x)=-\sum _{a,b}T^{ab}_A(x)N^{ab} \, ,
\label{T_A(x)}
\end{align}
\begin{align}
J=\int \!\! d^3{ \bf p} \sum _{h=\pm 1,a} h a^{a\dagger} _h({ \bf p}) a^a _h({ \bf p}) \, ,
\end{align}
and also find
\begin{align}
J\Ket{0}=T_A(x)\Ket{0}=0\,.
\end{align}

Furthermore, in a similar way to section II B, we can introduce two unphysical modes which transform as follows.
\begin{align}
[T_1(x),a^a _h({ \bf p})]= T_1^{ab}(x)a^b _h ({\bf p})+[a^a _h ({\bf p}),T_1^{bc}(x)]\tilde{N}^{bc}-a^a _S ({ \bf p})\,,
\end{align}
\begin{align}
[T_2(x),a^a _h({ \bf p})]= T_2^{ab}(x)a^b _h ({\bf p})+[a^a _h ({\bf p}),T_2^{bc}(x)]\tilde{N}^{bc}+iha^a _S({ \bf p})\,,
\end{align}
\begin{align}
[T_1(x),a _L({ \bf p})]=T_1^{ab}(x)a^b_L ({\bf p})-\sum_{h=\pm 1}a^a _h ({ \bf p})\,,
\quad
[T_2(x),a _L({ \bf p})]=T_2^{ab}(x)a^b_L ({\bf p})-i\sum_{h=\pm 1}ha^a _h ({ \bf p})\,,
\end{align}
\begin{align}
[T_1(x),a _S({ \bf p})]=T_1^{ab}(x)a^b_S ({\bf p})\,,
\quad
[T_2(x),a _S({ \bf p})]=T_2^{ab}(x)a^b_S ({\bf p})\,,
\end{align}
where,
\begin{align}
\tilde{N}^{ab}:=\int \!\! d^3{\bf p} \Lt\{\sum _{h=\pm 1} \!\!  a^{a\dagger}_h ({\bf p}) a^b_h ({\bf p}) 
+a^{a\dagger}_L ({\bf p}) a^b_S ({\bf p})+a^{a\dagger}_S ({\bf p}) a^b_L ({\bf p})\Rt\}\, .
\end{align}
Consequently, we obtain the local Lorentz transformation as follows.
\begin{align}
\tilde{U}(\Lambda ,x)A^\mu (x)\tilde{U}^{-1} (\Lambda ,x)
= (\Lambda ^{-1})^\mu _{\ \nu}\Lt(A^\nu (\Lambda x)+gf^{abc}A^{\nu b}(\Lambda x)\Theta ^c (\Lambda x,\alpha _1,\alpha _2)\Rt)\, .
\label{Lt tilde-Aa}
\end{align}

It seems that (\ref{Lt tilde-Aa}) is global gauge transformation.
However, we cannot regard there as so
because we can find the transformation at other point $y$ as
\begin{multline}
\tilde{U}(\Lambda ,y)A^\mu (x)\tilde{U}^{-1} (\Lambda ,y)
= (\Lambda ^{-1})^\mu _{\ \nu}(A^\nu (\Lambda x)+gf^{abc}A^{\nu b}(\Lambda x)\Theta ^c (\Lambda y,\alpha _1,\alpha _2)
\\
+gf^{abc}\tilde{N}^{bc}F(\Lambda (x-y)) )\, ,
\label{Lt tilde-Aa y}
\end{multline}
\begin{align}
F(x-y):=\int \!\! \f{d^3{\bf p}}{(2\pi )^3 2E_{{\bf p}}} \Lt( \alpha_1 \omega ^\mu ({\bf p}) \e ^{-ip(x-y)}
+\alpha_2 \tilde{\omega } ^\mu ({\bf p}) \e ^{ip(x-y)}\Rt)\, ,
\end{align}
namely, the proportional term to $\tilde{N}^{bc}$ is remaining.

\section{In other dimensions}
In this section, we show that gauge transformation is realized as representation of the little group in any spacetime dimensions.

We can obtain the Casimir operators of the $D(\geq 3)$-dimensional Poincar\'{e} algebra by replacing
\begin{align}
W_\mu\rightarrow  W_{\mu_1 \cdots \mu_{D-3}}:=-\f{1}{2}\varepsilon _{\mu_1 \cdots \mu_{D-3} \nu \rho \sigma}P^\nu M^{\rho \sigma}\, .
\end{align}
Then, we can find that the little group of massless particle is the $(D-2)$-dimensional Euclidean group $ISO(D-2)$;
\begin{align}
[T_A ,T_B ]=0\, ,\quad
[J_{AB},J_C ]=i(\delta _{BC}P_A -\delta _{AC}P_B )\,,
\n
\end{align}
\begin{align}
[J_{AB},J_{CD}]=-i(\delta  _{AC}J_{BD}+\delta  _{BD}J_{AC}-\delta  _{AD}J_{BC}-\delta  _{BC}J_{AD})\, .
\label{ISO}
\end{align}

\subsection{The 3-dimensions}
The little group is $ISO(1)$, namely, ${\mathbb R}$.
Accordingly, there is just a one continuous degree of freedom. 
In other words, there is no degree of freedom corresponding to helicity.
We define the translation generator of ${\mathbb R}$ and the polarization vector as follows.
\begin{align}
(T)^\mu _{\ \nu} =
\begin{pmatrix}
0 & -i  & 0 \\
-i & 0  & i \\
0 & -i  & 0
\end{pmatrix}\,,
\quad
\epsilon ^\mu :=(0,1,0) \,.
\end{align}
It is obvious that we can obtain this matrix and polarization vector from the case of 4-dimension by ignoring the y-component .
Therefore, we can find representation of the little group ${\mathbb R}$ by removing $T_2$ and $J$ from the case of 4-dimension.
Consequently, gauge transformation is also realized as representation of the little group in the 3-dimensional spacetime.

\subsection{The 5-dimensions}
The little group is $ISO(3)$.
Accordingly, there are three continuous degrees of freedom.
First, we define six generators of $ISO(3)$ acting on the Lorentz vectors as follows.
\begin{align}
(T_1)^\mu _{\ \nu} =
\begin{pmatrix}
0 & -i  &0 &0 & 0 \\
-i & 0  &0 &0 & i \\
0 & 0  &0 &0 & 0 \\
0 & 0  &0 &0 & 0 \\
0 & -i  &0 &0 & 0
\end{pmatrix}\,,
\quad
(T_2)^\mu _{\ \nu} =
\begin{pmatrix}
0 & 0  &-i &0 & 0 \\
0 & 0  &0 &0 & 0 \\
-i & 0  &0 &0 & i \\
0 & 0  &0 &0 & 0 \\
0 & 0  &-i &0 & 0
\end{pmatrix}\,,
\quad
(T_3)^\mu _{\ \nu} =
\begin{pmatrix}
0 & 0  &0 &-i & 0 \\
0 & 0  &0 &0 & 0 \\
0 & 0  &0 &0 & 0 \\
-i & 0  &0 &0 & i \\
0 & 0  &0 & -i & 0
\end{pmatrix}\,,
\end{align}
\begin{align}
(J_\pm)^\mu _{\ \nu}
:=\f{1}{\sqrt{2}}(J_{12}\pm iJ_{23})^\mu _{\ \nu}=\f{1}{\sqrt{2}}
\begin{pmatrix}
0 & 0 & 0 & 0 & 0 \\
0 & 0 & i & 0 & 0 \\
0 & -i & 0 & \mp 1 & 0 \\
0 & 0 & \pm 1 & 0 & 0\\
0 & 0 & 0 & 0 & 0
\end{pmatrix}\,,
\quad
(J)^\mu _{\ \nu}
:=(J_{31})^\mu _{\ \nu} =
\begin{pmatrix}
0 & 0 & 0 & 0 & 0 \\
0 & 0 & 0 & -i & 0 \\
0 & 0 & 0 & 0 & 0 \\
0 & i & 0 & 0 & 0\\
0 & 0 & 0 & 0 & 0
\end{pmatrix}\,,
\end{align}
Then, we define three polarization vectors as eigenvectors of $J$;
\begin{align}
\epsilon_{\pm 1} ^\mu :=(0,1,0,\pm i, 0)\, ,
\quad
\epsilon_0 ^\mu := (J^\dagger _\pm )^\mu _{\ \nu} \epsilon_{\mp 1} ^\nu =(0,0,-i,0, 0)\, .
\end{align}
 Therefore, we obtain equations similar to (\ref{epsilon(p)}) as follows.
\begin{align}
(1\mp i\theta )\epsilon_{\pm 1} ^\mu ({\bf p})-i\eta  _{\pm}\epsilon_0  ^\mu ({\bf p})
\simeq (\Lambda^{-1})^\mu _\nu \epsilon_{\pm 1} ^\mu ({\bf \Lambda p})+\f{\alpha_1 \pm i\alpha_3}{\sqrt{2}k}p^\mu \, ,
\label{epsilon(p)5D}
\end{align}
\begin{align}
\epsilon_0 ^\mu ({\bf p})-i(\eta  _- \epsilon_ {+1} ^\mu ({\bf p})+\eta  _+ \epsilon_ {-1} ^\mu ({\bf p}))
\simeq (\Lambda^{-1})^\mu _\nu \epsilon_0 ^\mu ({\bf \Lambda p})-\f{i \alpha_2}{k}p^\mu \, .
\end{align}
Now, $\eta  _\pm$ are each parameters of transformations of $J_{\pm}$ .

After that, we define actions of $SO(3)$ generators $J$ and $J_{\pm}$ on creation and annihilation operators as follows.
\begin{align}
[J,a_{\pm} ({\bf p})]=\mp a_{\pm} ({\bf p})\, ,
\quad
[J_+,a _{+1} ({\bf p})]=[J_-,a _{-1} ({\bf p})]=0\, ,
\quad
[J_+,a _{-1} ({\bf p})]=[J_-,a _{+1} ({\bf p})]=-a _0 ({\bf p})\, ,
\end{align}
\begin{align}
[J,a^\dagger _{\pm} ({\bf p})]=\pm a^\dagger _{\pm} ({\bf p})\, ,
\quad
[J^\dagger _+,a^\dagger _{+1} ({\bf p})]=[J^\dagger _-,a^\dagger _{-1} ({\bf p})]=0\, ,
\quad
[J^\dagger _+,a^\dagger _{-1} ({\bf p})]=[J^\dagger _-,a^\dagger _{+1} ({\bf p})]=a^\dagger _0 ({\bf p})\, ,
\end{align}
\begin{align}
[J,a_0 ({\bf p})]=[J,a^\dagger _0({\bf p})]=0\, ,
\quad
[J_{\pm},a _0 ({\bf p})]=-a _{\pm 1} ({\bf p})\, ,
\quad
[J^\dagger _{\pm},a^\dagger _0 ({\bf p})]=a^\dagger _{\pm 1} ({\bf p})\, .
\label{[j,a_0]}
\end{align}
By using (\ref{epsilon(p)5D}) to (\ref{[j,a_0]}), we obtain the Lorentz transformation of vector field similar to (\ref{Lt-A}).
However, the parameter of transformation is different from (\ref{Theta});
\begin{multline}
\Theta  (x,\alpha _1,\alpha _2,\alpha _3):=\f{i}{\sqrt{2}k}\int \f{d^4{ \bf p}}{(2\pi )^2\sqrt{2E_{{\bf p}}}}
\Lt[\alpha _1 \sum_{h=\pm 1}( a_h({\bf p}) \e ^{-ipx}-a^\dagger _h({\bf p})\e ^{ipx})
\Rt.\\ \Lt.
-\sqrt{2}i( a_0({\bf p}) \e ^{-ipx}+a^\dagger _0({\bf p})\e ^{ipx})
+ih\alpha _3\sum_{h=\pm 1}( a_h({\bf p}) \e ^{-ipx}+a^\dagger _h({\bf p})\e ^{ipx})\Rt]\, .
\end{multline}
Hence, we define three translation generators similar to (\ref{T_A(x)}),
but now,
\begin{align}
T_1^{ab}(x):=f^{abc}\f{g}{\sqrt{2}k}\int \f{d^4{ \bf p}}{(2\pi )^2\sqrt{2E_{{ \bf p}}}}\sum_{h=\pm 1}
\Lt[a^c_h({\bf p}) \e ^{-ipx}-a^{c\dagger} _h({\bf p}) \e ^{ipx}\Rt] \, ,
\end{align}
\begin{align}
T_2^{ab}(x):=-f^{abc}\f{ig}{k}\int \f{d^4{ \bf p}}{(2\pi )^2\sqrt{2E_{{ \bf p}}}}
\Lt[a^c_0({\bf p}) \e ^{-ipx}+a^{c\dagger} _0({\bf p}) \e ^{ipx}\Rt] \, ,
\end{align}
\begin{align}
T_3^{ab}(x):=f^{abc}\f{ig}{\sqrt{2}k}\int \f{d^4{ \bf p}}{(2\pi )^2\sqrt{2E_{{ \bf p}}}}\sum_{h=\pm 1}h
\Lt[a^c_h({\bf p}) \e ^{-ipx}+a^{c\dagger} _h({\bf p}) \e ^{ipx}\Rt] \, ,
\end{align}
\begin{align}
N^{ab}:=\int \!\! d^3{\bf p} \!\!  \sum _{h=\pm 1,0} \!\!  a^{a\dagger}_h ({\bf p}) a^b_h ({\bf p})\, .
\end{align}
In the case of non-abelian extension, then, we can find that these operators satisfy $ISO(3)$ algebra;
\begin{align}
[J,T_1(x)]=iT_3(x)\,,
\quad
[J,T_3(x)]=-iT_1(x)\,,
\quad
[J,T_2(x)]=[T_A(x),T_B(x)]=0\,,
\n
\end{align}
\begin{align}
[J_{\pm},T_1(x)]=-\f{i}{\sqrt{2}}T_2(x)\,,
\quad
[J_{\pm},T_2(x)]=\f{1}{\sqrt{2}}\Lt( iT_1(x)\pm T_2(x)\Rt)\,,
\quad
[J_{\pm},T_3(x)]=\mp \f{1}{\sqrt{2}}T_2(x)\,,
\end{align}  
and generate non-abelian gauge transformation similar to (\ref{Lt-Aa}) after all.

\subsection{Even-dimensions}
We consider the case of $D=2N+N$.
So, the little group is $ISO(2N)$.
We define generators of $ISO(2N)$ acting on the Lorentz vectors as follows.
\begin{align}
(T_A)^\mu _{\ \nu} =
\begin{pmatrix}
0 & -i\delta_{A\nu}  & 0 \\
-i\delta_{A\mu} & 0  & i\delta_{A\mu} \\
0 & -i\delta_{A\nu}  & 0
\end{pmatrix}\,,
\quad
(J_{AB})^\mu _{\ \nu} =
\begin{pmatrix}
0 & 0  & 0 \\
0 & i(\delta_{A\mu}\delta_{B\nu}-\delta_{B\mu}\delta_{A\nu})  & 0 \\
0 & 0  & 0
\end{pmatrix}\,.
\label{ISO(N)matrix}
\end{align}
There are the $N$ Cartan generators of $SO(2N)$ which are $2\times2$ block diagonalized matrices; $H_n:=J_{2n-1,2n}$\cite{Georgi}.
Therefore, we introduce $2N$ polarization vectors as the eigenvectors of $H_n$ similar to 4-dimensional case, namely, labeled by helicity $h=\pm 1$ for each $n$.
Hence, if we only consider $H_n$ about subalgebra of $SO(2N)$, we can obtain representation of $ISO(2N)$ as $n$ copies of $ISO(2)$ 
because each translation generators corresponding to two directions of $H_n$ are commutable to the other Cartan generators,
Then, the actions of other generators of $SO(2N)$ are exchanging each $n$ states.
Thus, we can also find the representation of these generators similar to the case of $ISO(2)$ with redefining the Cartan generators.
Consequently, we can reproduce gauge transformation form Lorentz transformation by similar way to the 4-dimensional case, although we do not show this explicitly.
 
\subsection{Odd-dimensions}
We consider the case of $D=2N+3$ and the little group $ISO(2N+1)$.
We define generators of $ISO(2N+1)$ similar to (\ref{ISO(N)matrix}).
In contrast, there are $N+1$ Cartan generators of $SO(2N+1)$ which are $N-1$, $2\times2$ and one $3\times3$ block diagonalized matrices.
Therefore, we introduce polarization vectors which are $2N-2$ vectors as the even-dimensional case and three vectors as the 5-dimensional case
and then we can reproduce gauge transformation form Lorentz transformation by similar way to the even-dimensional case.

\section*{SUMMARY}
In this paper we investigated the relations between gauge symmetry and the Lorentz symmetry, specifically the symmetry of little group.
A Free single-particle state was classified by mass, spatial momentum, and representation of the little group.
In particular, the little group of massless particles in $D$-dimensional spacetime is $ISO(D-2)$, namely $D-2$-dimensional Euclidean group
which contains rotations and translations.
Therefore, these states have continuous degree of freedom but there are unphysical.
Hence, we have assumed that the action of translations are trivial.
As a result, we realized abelian gauge transformation form the Lorentz transformation of the massless vector, also tensor, field because of the action of little group translations on polarization vectors.
Moreover,  by including the unphysical modes, we constructed the representation of little group as transforming physical mode to unphysical modes, and vice versa
and we realized ordinary Lorentz transformation.
Furthermore, we obtained the same physical state condition with the Nakanishi-Lautrup formalism for unitarity.
After that, we extended these results to the case of non-abelian gauge transformation
as Lorentz transformation was restricted to the local transformation
and the representation of little group was realized as tensor operators acting on gauge group constructed from creation and annihilation operators.
Finally, we showed that we can obtain gauge transformation by similar way in any spacetime dimensions.

We obtained non-abelian gauge transformation form Lorentz transformation of physical single-particle stats.
Consequently, if we impose the Lorentz symmetry for Lagrangian of vector field, it should have also gauge symmetry.
So, we conclude it is not necessarily to impose gauge symmetry explicitly.
It is introduced automatically if we impose only the Lorentz symmetry.

Finally, we comment that the extension to curved backgrounds and general coordinate transformation.
It is easy to extend the procedure of this paper to curved backgrounds by using vielbein, because we can regard the Lorentz transformation as local one. 
Then, by using connection 1-form field, we can consider whether nonlinear general coordinate transformation is reproduced by our procedure.
Furthermore, as generalization of the case of connection 1-form, we can consider interacting higher-spin gauge theory.
It is known to construct interacting higher-spin gauge theory is difficult, but it could be found by our procedure.

\section*{Acknowledgments.}

The author is grateful to Prof. Taichiro~Kugo, Prof. Tadakatsu~Sakai and Prof. Shin'ichi~Nojiri for discussions,
and supported in part by Global COE Program of Nagoya University (G07)
provided by the Ministry of Education, Culture, Sports, Science \& Technology.

\appendix*
    \renewcommand{\theequation}{A.\arabic{equation} }
    \setcounter{equation}{0} 
    
\section{``No-go theorem'' for higher-spin fields with non-abelian gauge symmetry}
In section III, we have shown that non-abelian gauge transformation of spin-1 field are generated by translations of the little group $ISO(N)$.
Thus, we can conjecture that for any integer spin field, generally, gauge transformation should be realized as Lorentz transformation.
Indeed, as shown in section II A, abelian gauge transformation for any integer spin fields are identified actions of the little group.
Therefore, we consider to construct non-abelian gauge transformation of more then spin-2 fields similarly to the case of abelian transformation.
In this section, we consider only the 4-dimensional case for simplicity.

We introduce spin-n fields with indices of adjoint representation of $G$ from (\ref{HS}) by replacing $a_{nh}({\bf p})\rightarrow a^a_{nh}({\bf p})$;
\begin{align}
\phi ^{\mu_1\cdots\mu_n a} (x):=
\int \f{d^3{ \bf p}}{(2\pi )^\f{3}{2}\sqrt{2E_{{\bf p}}}}\sum_{h=\pm 1} \Lt[ \epsilon ^{\mu_1\cdots\mu_n} _{h} ({\bf p}) a^a_{nh}({\bf p}) \e ^{-ipx}
+\epsilon ^{*\mu_1\cdots\mu_n} _{h} ({\bf p}) a^{a\dagger} _{nh}({\bf p}) \e ^{ipx} \Rt] \,,
\end{align}
and consider trivial generalization of non-abelian gauge transformation (\ref{Lt-Aa}) as follows.
\begin{multline}
U(\Lambda )\phi ^{\mu_1\cdots\mu_n a} (x)U^\dagger (\Lambda )
=(\Lambda ^{-1})^{\mu_1} _{\ \nu_1}\cdots(\Lambda ^{-1})^{\mu_n} _{\ \nu_n}
\Lt(\phi ^{\nu_1\cdots\nu_n a}  (\Lambda x) +\p ^{(\nu_1} \xi  ^{\nu_2\cdots\nu_n)a}   (\Lambda x)
\Rt.\\ \Lt.
-gf^{abc}\Theta ^{(n)b} (\Lambda x,\alpha _1,\alpha _2) \phi ^{\nu_1\cdots\nu_n c}(\Lambda x)
\Rt)\, ,
\label{Lt-HSa}
\end{multline}
where,
\begin{align}
\Theta ^{(n)a}  (x,\alpha _1,\alpha _2):=\f{i}{\sqrt{2}k}\int \f{d^3{ \bf p}}{(2\pi )^\f{3}{2}\sqrt{2E_{{\bf p}}}}\sum_{h=\pm 1}
\Lt[\alpha _1 ( a^a_{nh}({\bf p}) \e ^{-ipx}-a^{a\dagger} _{nh}({\bf p})\e ^{ipx}) +ih\alpha _2( a^a_{nh}({\bf p}) \e ^{-ipx}+a^{a\dagger} _{nh}({\bf p})\e ^{ipx})\Rt]\, .
\end{align}
It is easy to reproduce the transformation (\ref{Lt-HSa}) similar to the spin-1 case by also replacing $a_{nh}({\bf p})\rightarrow a^a_{nh}({\bf p})$ for $J$ and $T_A(x)$.
However, such generators are not representation of $ISO(2)$;
\begin{align}
[J,T_1(x)]=inT_2(x)\, ,
\quad
[J,T_2(x)]=-inT_1(x)\, .
\end{align}
In other words, the transformation (\ref{Lt-HSa}) is not realized as the Lorentz transformation.
Therefore, if the conjecture; {\it gauge transformation should be realized as Lorentz transformation} is true,
we can conclude that there is no higher-spin field with (internal) non-abelian gauge symmetry in flat spacetime.
This conclusion is consistent with Coleman-Mandula theorem\cite{Coleman-Mandula}.

\end{document}